\documentclass[prl,twocolumn,showpacs]{revtex4}

\usepackage{graphicx}

\begin{document}

\title{Melting of Branched RNA Molecules}
\author{Ralf Bundschuh}
\affiliation{Department of Physics, Ohio State University, Columbus,
OH 432110-1106}
\author{Robijn Bruinsma}
\affiliation{The University of California at Los Angeles, Los Angeles, CA 90049}

\begin{abstract}
Stability of the branching structure of an RNA molecule is an
important condition for its function. In this letter we show that the
melting thermodynamics of RNA molecules is very sensitive to their
branching geometry for the case of a molecule whose groundstate has
the branching geometry of a Cayley Tree and whose pairing interactions
are described by the G\=o model. Whereas RNA molecules with a linear
geometry melt via a conventional continuous phase transition with
classical exponents, molecules with a Cayley Tree geometry are found
to have a free energy that {\em seems} smooth, at least within our
precision. Yet, we show analytically that this free energy in fact has
a mathematical singularity at the stability limit of the ordered
structure. The correlation length appears to diverge on the
high-temperature side of this singularity.
\end{abstract}


\maketitle


A fundamental principle of statistical mechanics states that phase
transitions are not possible for one-dimensional systems unless
long-range interactions are present. It thus came as a surprise when
Poland and Scheraga (PS) showed~\cite{ps} that an infinite, linear
molecule composed of two flexible polymer strands bound together by a
local attractive interaction does undergo a true phase transition at
the temperature where the two strands separate. The required
long-range correlations are due to the fact that the partition
function of a strand separation ``bubble'' has a power-law dependence
on size. The mean bubble size, the correlation length, diverges at the
critical point if the transition is
continuous~\cite{ordercomment}. This observation was particularly
interesting because that system could be viewed as a simple model for
the denaturation of double-stranded B-DNA molecules.

The PS mechanism can be extended to the melting of --- more complex
--- RNA molecules~\cite{Alberts}. In a biological context, RNA
molecules usually operate in a {\em single-stranded} mode. This single
strand can however bend onto itself so the bases of the strand can
self-pair into a pattern of bubbles and ``stems'' that can be
displayed in the form of a tree-like planar graph, the ``{\em
  secondary structure}''~\cite{higgsreview}.  The minimum-energy
secondary structure of a functional RNA molecule plays an important
role in its functioning, and can be predicted from the primary
sequence of nucleotides~\cite{Zuker}. Melting of a minimum-energy
secondary structure produces a ``molten globule'' state with the
molecule fluctuating over a range of different secondary
structures~\cite{higgsmolten}. Importantly, in this molten globule
state, most bases remain paired in contrast to the fully denatured
state, which is favored at higher temperatures, with most of bases
unpaired. In his pioneering paper of 1968~\cite{deGennes}, de Gennes
showed that the partition function $G(L)$ of a large RNA molecule
fluctuating over {\em all} possible secondary structures with
identical non-specific pairing energies has a power-law dependence on
size of the form $\frac{z_0^L}{L^\theta}$ with $\theta=3/2$. Bundschuh
and Hwa~\cite{bh} (BH) extended this result to show that if the
groundstate of an RNA molecule is a long, linear hairpin stabilized by
specific pairing energies then thermal fluctuations in the form of
molten-globule bubbles produce a melting thermodynamics that,
formally, has the same form as that of the PS model.

\begin{figure}
\begin{center}
\includegraphics[width=0.7\columnwidth]{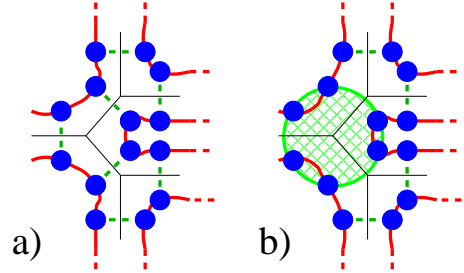}
\caption{Single-stranded RNA molecule having a branched secondary structure
that follows the outline of a Cayley Tree. Nucleotides are
schematically indicated by circles, bonds between nucleotides by a
solid line and complementary pairing interactions by dashed
lines. a) Groundstate structure with pairing restricted to a
complementary ``native'' pair for each branch of
the Cayley Tree. b) In a molten globule bubble (hatched) all
possible pairing interactions are permitted.}
\end{center}
\vspace*{-5mm}
\end{figure}

Actual RNA secondary structures have a branched, tree-like form, which
raises the question how and if the melting thermodynamics of such a
form differs from that of a simple hairpin. It is the experience with
many statistical mechanics models defined on tree-like geometries
without circuits that they exhibit {\em mean-field} type critical
behavior. Since, in the absence of excluded volume interactions, the
critical properties of the PS and BH models {\em already} are of
mean-field character, one would expect the free energy of branched
secondary structures to exhibit mean-field critical behavior. In this
letter we will show that in fact the melting thermodynamics of a
particular, highly branched secondary structure is highly anomalous:
the numerically computed free energy {\em appears} to have no
singularity, yet, surprisingly, we can demonstrate analytically that
the free energy {\em does} have a mathematical singularity at a point
where the branched groundstate becomes unstable. The correlation
length appears to diverge on the high-temperature side of the
singularity, yet, on the low-temperature side this singularity is not
associated with a divergence of the correlation length.

To demonstrate these claims, we consider an RNA molecule that has the
shape of a {\em Cayley Tree} (see Fig.1). In the groundstate, the
single strand traces out the perimeter of the tree, starting and
ending at the root of the tree, with each branch of the tree occupied
by a single complementary base-pair. The size of the molecule is
indexed by the level $k$ of the tree that is related to the total
sequence length $N(k)$ of the strand by $N(k) = 2^{k+2}-2$ bases (a
$k = 1$ tree is here a three-armed star with one base-pair per
arm). After sequentially numbering the bases of the strand, one can
denote this ``designed'' groundstate by a list $S =\{i_1, j_1\},\{i_2,
j_2 \},\ldots, \{i_M , j_M \}$ of complementary pairs. We will assign
a specific binding energy $-\widetilde{\varepsilon}$ to any pair in
this list. Pairing between two bases that do not appear in this list
still will be allowed as long as it does not introduce any circuits
(or ``pseudoknots'') but the associated binding energy $-\varepsilon$
will be assumed to be less attractive than
$-\widetilde{\varepsilon}$. This definition of the pairing energy,
known as a ``G\=o Model''~\cite{go}, guarantees that the secondary
structure of the groundstate has the shape of a Cayley Tree.

Our strategy to obtain the finite-temperature partition function of
the system is to generalize the method of BH for the one-dimensional
case by expressing the partition function in the form of a sum over
all possible insertions of molten globule bubbles in the ground-state
structure. Inserting a bubble into a Cayley Tree is more complex than
into a linear structure: a bubble inside the tree can have different
numbers of branches attached to it so one has to keep track of
different bubble species. We will show elsewhere that, within the G\=o
model, the partition function of any ``designed'' secondary structure
can be written as a sum over configurations classified according to
the size $2n$ of the ``accessible'' open bubble located at the base of the
tree (see Fig.1). Here, $n$ is the number of base pairs of the open
bubble. Specifically, the partition function $Z(k)$ can be written as:
\begin{equation}\label{eq_ZthroughW}
Z(k)=\sum_{n=0}^{N(k)/2}G(2n)W(k,n).
\end{equation}
In Eq.~(\ref{eq_ZthroughW}), $G(M)\approx\frac{z_0^M}{M^\theta}$ is the
partition function of a strand of length $M$ with no specific pairing,
i.e. all paired bases have a binding energy $-\varepsilon$ {\em even
if the pair appears in the list $S$ of specific groundstate
pairs}. Next, $W(k,n)$ is a {\em restricted} partition function, i.e.,
the partition function of a molecule with $n$ accessible bases in the
open bubble at the root, but {\em not} including the configurations of
the open bubble. This restricted partition function can be written as
a sum over all possible bubble insertions:
\begin{equation}\label{eq_Winloops}
W (k, n)=\sum_{S'(n)\subset S} (\widetilde{q}− q)^{|S'(n)|}
\prod_{\{L(S')\}}G(L(S')).
\end{equation}
Here, $q = \exp(−\beta\varepsilon)$ and $\widetilde{q}= \exp
(−\beta\widetilde{\varepsilon})$ while $S'(n)$ is any of the subsets
of $S$ that is compatible with $n$ base-pairs in the open bubble at
the root. The number of specifically paired bases of $S'(n)$ is
denoted by $|S'(n)|$. Each term of Eq.~(\ref{eq_Winloops}) represents
a secondary structure having $|S'(n)|$ specifically paired bases
linked together by a distribution of closed bubbles with sizes
$\{L(S')\}$.

Using Eq.~(\ref{eq_Winloops}) one can construct two linked recursion
relations. First, cut a tree with restricted partition function
$W(k,n)$ into two equal sized sub-trees with level index $k-1$. The
number of accessible base pairs of the two sub-trees together must add
to $n-1$, as we removed one pair by the cutting operation. Because we
permit no circuits, the restricted partition function of a level $k$
tree and $n>0$ can be expressed in terms of a product of the
restricted partition functions of two $k-1$ level sub-trees:
\begin{equation}\label{eq_recngt0}
W(k,n)=\sum_{m=0}^{n-1}W(k-1,m)W(k-1,n-1-m)
\end{equation}
with $W(k-1, m) = 0$ if $m > 2^{k}-1$. The $n=0$ case --- a tree with
no bubble at the root --- must be treated separately. Take the first
complementary pair at the root of the tree out of the partition
function, and then sum over all possible sizes for the bubble that
immediately follows this pair (including a bubble of zero size). Now
treat {\em that} bubble as the bubble at the root of a new tree that can
again be cut into two equal parts in the same way as before. This
leads to a second recursion relation:
\begin{eqnarray}
\lefteqn{W(k,n)=}\hspace*{3mm}\label{eq_recneq0}\\
&&(\widetilde{q}\!-\!q)\!\!\!\sum_{n_1=0}^{2^k\!-\!1}
\sum_{n_2=0}^{2^k\!-\!1}\!\!
W(k\!-\!1,n_1)W(k\!-\!1,n_2)G(2(n_1\!+\!n_2))\nonumber
\end{eqnarray}
Equations~(\ref{eq_recngt0}) and~(\ref{eq_recneq0}) together
constitute a complete set of recursion relations for $W(k,n)$ that can
be solved iteratively. The initial conditions for the recursion
relations are
$W(1,0)=(\widetilde{q}-q)[1+4q+q^2+2\widetilde{q}+\widetilde{q}^2]$,
$W(1,1)=(\widetilde{q}-q)^2$, $W(1,2)=2(\widetilde{q}-q)$, and
$W(1,3)=1$, as follows by inspection.

We carried out this iteration procedure numerically up to level $k=19$
for different values of
$\widetilde{q}=\exp(\beta\widetilde{\varepsilon})$ and for fixed
$q=4$. In Fig.~2 we show the second derivative of the free energy per
site with respect to $\widetilde{q}$, which effectively correspond to
the heat capacity. As one increases the value of $k$, a maximum
develops near $\widetilde{q}$. However, within the numerical
precision, the free energy per site does {\em not} develop a
thermodynamic singularity in the large $N$ limit. This must be
contrasted with the case where the molecule has a linear hairpin
groundstate, in which case the heat capacity very clearly develops
such a singularity for much smaller system sizes (see inset of Fig.2).

\begin{figure}[tbh]
\begin{center}
\includegraphics[width=0.8\columnwidth]{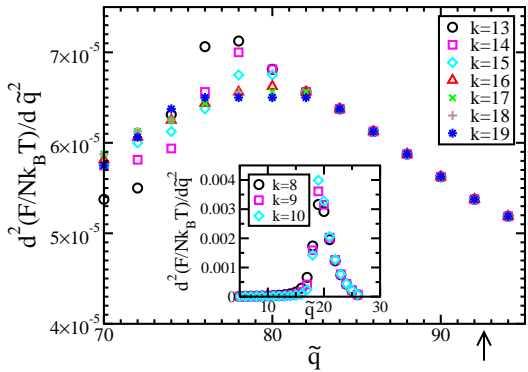}
\caption{Second derivative of the free energy with respect to the
Boltzmann weight \protect$\widetilde{q}$ of specifically paired bases
plotted as a function of \protect$\widetilde{q}$ for different values
of the level \protect$k$ of the Cayley tree groundstate. The free
energy was computed numerically from the recursion relations
Eqs.~(\ref{eq_recngt0}) and~(\ref{eq_recneq0}) and expressed in units
of $N k_BT$ with $N$ the sequence length of the RNA strand. The arrow
denotes the location of the mathematical singularity associated with
melting of the root of the Cayley Tree. Inset: same except that the
groundstate is a linear hairpin groundstate. A singularity develops
near \protect$\widetilde{q}=18.4$.}
\end{center}
\vspace*{-5mm}
\end{figure}

In order to examine {\em sub-leading} contributions to the free
energy, i.e., terms that are small compared to the leading term
proportional to $N$, we also computed the ``pinching free
energy''
\begin{equation}\label{eq_defpinchfe}
\Delta F(k)/k_B T\equiv \ln Z(k+1)-2\ln Z(k)
\end{equation}
For example, in a molten-globule phase the partition function should
have the asymptotic scaling form $a^+z_0^N/N^{3/2}$ for large $N$. The
pinching free energy $\Delta F(k)/k_BT=\frac{3}{2}(k+2)\ln2-\ln a^+$
then would have a {\em linear} dependence on $k$, with slope $3/2$. In
an ordered phase, the partition function should scale as $a^-z_0^N$
for large $N$, in which case $\Delta F(k)/k_BT=-\ln a^-$ should be a
constant independent of $k$. Figure~3 shows that, for $\widetilde{q}$
values up to $80$, $\Delta F(k)$ indeed has a linear dependence on $k$
for large $k$, with a slope close to $3/2\ln2$. This indicates that,
for $\widetilde{q}$ values below $80$, the tree is in the
molten-globule phase. Since for the corresponding case of a linear
groundstate, the melting point is as low as $\widetilde{q}_c\!=\!18.4$ for
$q\!=\!4.0$, we are forced to conclude that branching has a powerful
{\em destabilizing} effect on the ordered state.

For smaller $k$ values, the pinching free energy is a constant, which
indicates that the ordered groundstate dominates over shorter length
scales. The crossover point between the two regimes can be interpreted as a
{\em correlation length} $\xi$ whose physical meaning would be that of
the typical size of smaller ordered Cayley Tree type structures
imbedded in a larger molten-globule state. The value of $\xi$
increases with $\widetilde{q}$ according to Fig.3 and beyond
$\widetilde{q}= 80$ it exceeds our maximum system size ($N=10^6$). A
fit to a power-law $\xi\sim(\widetilde{q}_c-\widetilde{q})^{-\nu}$
produces a correlation length exponent $\nu\approx2.1$ and a critical
$\widetilde{q}_c\approx80$.

\begin{figure}[tbh]
\begin{center}
\includegraphics[width=0.8\columnwidth]{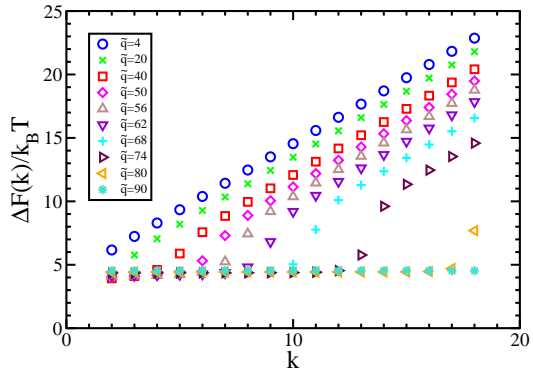}
\caption{Numerically computed ``pinching'' free energy $\Delta F(k)$
(see Eq.~(\ref{eq_defpinchfe})) versus the level $k$ of the Cayley
Tree. For $\widetilde{q}$ larger then $90$, $\Delta F(k)/ k_BT$ is
independent of $k$, consistent with the ordered groundstate. For
$\widetilde{q}$ less than then $20$, $\Delta F(k)/k_B T$ can be fitted
by the relation $\Delta F(k)/ k_BT\approx(k + 2)\ln2- \ln a^+$ for the
molten globule state. The cross-over point between these two regimes
for intermediate values of $\widetilde{q}$ marks the size of the ordered,
correlated regions in the molten globule state. For $\widetilde{q}$ above
$80$, the size of the correlated regions exceeds the system size.}
\end{center}
\vspace*{-5mm}
\end{figure}

Can we really be sure that there {\em is} a thermodynamically stable,
ordered phase at low but finite temperatures or might the groundstate
only appear at $T=0$?  In the ordered phase, the restricted partition
function would be expected to scale as $W(N,n)\sim w(n)z_0^N$
asymptotically for large $N$. Here, $w(n)$ is the fraction of
configurations that have an open bubble at the root of size $n$. If we
insert this Ansatz into the recursion relation Eq.~(\ref{eq_recngt0}),
we obtain the following fixed-point condition:
\begin{equation}\label{eq_wrecurs}
w(n) = \sum_{m=0}^{n-1}w(m)w(n-1-m)
\end{equation}
This equation can be solved by applying the discrete Laplace Transform
$\widehat{w}(z)\equiv\sum_{m=0}^\infty w(m)z^{-m}$. The solution
$\widehat{z}=\frac{z}{2}-\sqrt{\frac{z^2}{4}-zw(0)}$ has a branch-cut
starting at $z=1/4w(0)$, with $w(0)$ an undetermined constant. After
applying an inverse Laplace Transform, one finds that $w(n)$ actually
has the same scaling form as the partition function of a molten
globule:
\begin{equation}\label{eq_wscale}
w(n)\approx\frac{\exp[-n\ln(1/4w(0))]}{n^{3/2}}
\end{equation}
However, the mathematical origin of the $n^{-3/2}$ factor is here a
combinatorial factor that reflects the different ways one can
partition the open bubble between the two sub-trees. We may interpret
$\xi\sim1/\ln(1/4w(0))$ as the characteristic size of a molten globule
bubble at the root of the tree in the ordered phase. Numerical
iteration of the recursion relations for $W(k,n)$ for
$\widetilde{q}=150$ and $q=4$ were found to be consistent with
Eq.~(\ref{eq_wscale}). If one uses $W (N, n)\approx w(n)z_0^N$ in the
remaining recursion relation Eq.~(\ref{eq_recneq0}), with
Eq.~(\ref{eq_wrecurs}), one obtains the following self-consistency
relation for the unknown $w(0)$:
\begin{equation}\label{eq_selfconsistency}
w(0)=\frac{\widetilde{q}-q}{2\pi i}\oint\frac{1}{z}\widehat{G}(z)
\widehat{w}(1/z)^2\mathrm{d}z
\end{equation}
Here, $\widehat{G}(z)$ is the discrete Laplace Transform of
$G(L)$~\cite{exactG}, which has a branch-cut that terminates at $z =(1
+ 2\sqrt{q})^2$. The integration contour in
Eq.~(\ref{eq_selfconsistency}) must run inside an annulus in the
complex plane that surrounds the origin passing the real axis {\em
outside} the branch-cut of $\widehat{G}(z)$ that terminates at $(1 +
2\sqrt{q})^2$ but {\em inside} the branch-cut of $\widehat{w}(1/z)$
that starts at $z = 1/4w(0)$ . That means that the contour integral
can only be carried out as long as
\begin{equation}\label{eq_conditiononw}
w(0)\le\frac{1}{4(1+2\sqrt{q})^2}
\end{equation}
The partition function develops a mathematical singularity when the
two branch cuts merge, i.e., when Eq.~(\ref{eq_conditiononw}) reduces
to an equality. At that point, the partition
$w(n)\approx\frac{[1/4w(0)]^n}{n^{3/2}}$ of the root bubble has the
same form as the partition
$G(n)\approx\frac{(1+2\sqrt{q})^{2n}}{n^{3/2}}$ for a molten globule
of the same size. We can identify $w(0)=\frac{1}{4(1+2\sqrt{q})^2}$ as
the stability limit of the groundstate. Note that the (low
temperature) correlation length $\xi\sim1/\ln(1/4w(0))$ {\em cannot}
diverge at the stability limit. The critical value $\widetilde{q}_c$
for $\widetilde{q}$ at the stability limited is now easily obtained by
noting that $w(0)$ is small compared to one. Expanding the argument of
the contour integral in powers of $w(0)$ leads to: 
\begin{equation}\label{eq_wexpansion}
(\widetilde{q}\!-\!q)^{-1}\!\approx\!
w(0)\!+\!2(1\!+\!q)w(0)^2\!+\!5(1\!+\!6q\!+\!2q^2)w(0)^3\!+\!\ldots
\end{equation}
If Eq.~(\ref{eq_wexpansion}) is combined with
$w(0)=\frac{1}{4(1+2\sqrt{q})^2}$ one finds that for $q = 4$, the
singularity is at $\widetilde{q}\approx92.6$.

Surprisingly, the numerically computed free energy per site shown in
Fig.~2 exhibits no singular dependence on $\widetilde{q}$ in that
range. This is not inconsistent because $w(n)$ only contributes a
sub-leading term to the total free energy. On the other hand, the
correlation length obtained from the pinching free energy appears to
diverge near $\widetilde{q}_c$. We encountered however strong {\em
finite-size effects} in the numerical solution of the recursion
relations for $\widetilde{q}$ values in the range between $80$ and
$90$ which make it difficult to numerically explore the critical
properties in more detail. In addition, over that range of
$\widetilde{q}$ values, our fixed-point scaling Ansatz appears not to
be valid, at least for $k$ values less than $19$. Instead, the reduced
partition function scales as $W (k, n)/ W (k,0)\!\approx\! N(k)g(n/ N(k))$
with $g(x)$ a scaling function that is nearly linear for small values
of~$x$.

In summary, a branched RNA molecule in the form of a Cayley Tree
undergoes a phase transition from the branched groundstate to a molten
globule phase if one reduces the energetic bias for the
groundstate. The stability of the branched groundstate against thermal
fluctuations is significantly less than that of the linear
groundstate. Branching does {\em not} produce mean-field critical
behavior but, instead, smears out the specific heat anomaly that
characterizes systems with a linear groundstate. On the
``high-temperature'' side of the melting transition, numerical
solution of the recursion relation produces a diverging correlation
length. We showed - analytically - that on the low temperature side
the designed groundstate becomes unstable at a critical point where
the free energy develops a mathematical singularity not associated
with a divergence of the correlation length.

Experimental studies comparing the melting characteristics of large,
branched RNA molecules with that of linear, unbranched molecules that
could probe this exotic form of melting have not yet been carried out
but such systems would be  fascinating laboratories for statistical
mechanics. An important question in this respect would be the role of
excluded volume interactions and of ``tertiary'' pairing interactions,
i.e., pairing interactions that introduce, for example,
pseudo-knots. Excluded volume interactions in general tend to suppress
thermal fluctuations and possibly could restore the thermodynamic
singularity in the free energy per site that was encountered for
linear molecules. Tertiary interaction could have the effect of
turning a branched, secondary template into a three dimensional
gel-like structure, in which case the transition to the molten-globule
state could resemble the melting transition of a bulk solid material.

{\bf Acknowledgements:} We would like to thank the Aspen Center for
Theoretical Physics for its hospitality. RB would like to acknowledge
support by the NSF under DMR Grant 0404507. This paper is dedicated to
the memory of Pierre-Gilles de Gennes.


\end{document}